\documentclass[english,twocolumn,aps,prl, superscriptaddress,showpacs, amsmath,floatfix,,nofootinbib]{revtex4-2}

\usepackage{color,xcolor}

\usepackage{natbib}
\usepackage[T1]{fontenc}
\usepackage{units}
\usepackage{amssymb,amsmath}
\usepackage{graphicx}
\usepackage{esint}
\usepackage{bm}
\usepackage{ulem}
\usepackage[colorlinks=true, citecolor=blue]{hyperref}
\setcitestyle{journalcolor=blue}

\begin{document}

\title{Nernst Sign-Reversal in the Hexatic Vortex Phase of Weakly  Disordered \\a-MoGe Thin Films}

\author{Y. Wu$^*$} 
\address{Department of Physics and Jack and Pearl Resnick Institute and Institute of Nanotechnology and Advanced Materials,
  Bar-Ilan University, Ramat-Gan 52900, Israel}
\author{A. Roy$^*$}
\address{Department of Physics and Jack and Pearl Resnick Institute and Institute of Nanotechnology and Advanced Materials,
  Bar-Ilan University, Ramat-Gan 52900, Israel}
\address{Department of Physics, Birla Institute of Technology and Science Pilani - K K Birla Goa Campus, Zuarinagar, Goa 403726, India}
  \author{S. Dutta}
\address{Tata Institute of Fundamental Research, Homi Bhabha Road, Mumbai 400005, India}
\author{J. Jesudasan}
\address{Tata Institute of Fundamental Research, Homi Bhabha Road, Mumbai 400005, India}
\author{P. Raychaudhuri}
\address{Tata Institute of Fundamental Research, Homi Bhabha Road, Mumbai 400005, India}
\author{A. Frydman}
\address{Department of Physics and Jack and Pearl Resnick Institute and Institute of Nanotechnology and Advanced Materials,
  Bar-Ilan University, Ramat-Gan 52900, Israel}

\date{\today}

\begin{abstract} 

The hexatic phase is an intermediate stage in the melting process of a 2D crystal due to topological defects. Recently, this exotic phase was experimentally identified in the vortex lattice of 2D weakly disordered superconducting MoGe by scanning tunneling microscopic measurements. Here we study this vortex state by the Nernst effect, which is an effective and sensitive tool to detect vortex motion, especially in the superconducting fluctuation regime. We find a surprising Nernst sign reversal at the melting transition of the hexatic phase. We propose that they are a consequence of vortex dislocations in the hexatic state which diffuse preferably from the cold to hot. 

\end{abstract}
\maketitle

\def\thefootnote{*}\footnotetext{Equal contribution}\def\thefootnote{\arabic{footnote}}

\begin{figure}[t]
\includegraphics[width=0.5\textwidth]{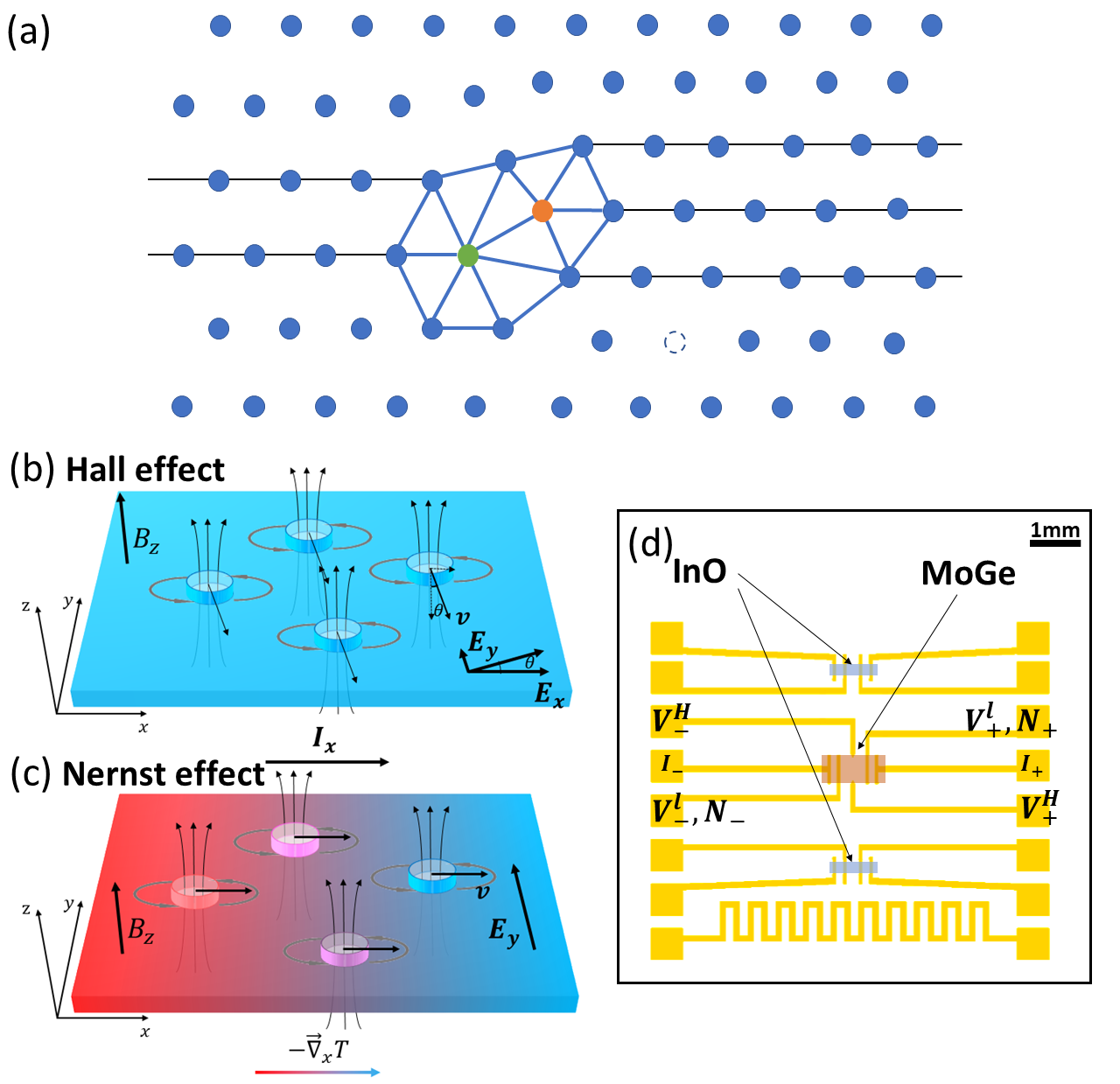}
\caption{(Color online)
Schematics of a dislocation (a) the vortex Hall effect (b), vortex Nernst effect (c) and sample configuration (d). (a) A dislocation in a hexagonal lattice is composed of two disclinations (5-fold (orange) and 7-fold (green) orientational symmetry). Note that a additional lattice line is present on the right side of the defect. The dashed circle is a vortex vacancy. (b) In the flux-flow regime of a superconducting film, a current $I_x$ generates a collective vortex velocity $\boldsymbol{v}$. The vortex motion produces an Electric field whose x component $E_x$ contributes to the dissipative longitudinal resistivity while y component $E_y$ gives rise to the Hall effect. The Hall angle $\theta$ is defined as $\tan{\theta}=\frac{E_y}{E_x}$. (c) The vortex Nernst effect is denoted as superconducting vortices moving in x direction with velocity x anti-paralleled with a thermal gradient $-\nabla_x{T}$, which generates a transverse electric field $E_y$. (d) A thin film of a-MoGe is pulsed laser deposited onto a chip of borosilicate glass. Electrodes are connected for passing excitation current ($I_+$, $I_-$) and measuring the longitudinal($V^l_+$, $V^l_-$), Hall($V^H_+$, $V^H_-$) and Nernst voltages ($N_+$, $N_-$). A meander made of 4nm thick Cr and 30nm thick Au and two films of insulating InO are fabricated for heating and the thermometry.}
\label{nernst}
\end{figure}

The melting process of a two-dimensional (2D) crystal has been a fundamental topic in condensed matter physicists research for decades. It is considered a topological phase transition that is mediated by topological defects, namely dislocations and disclinations. In two dimensions, a dislocation is a point defect which is a local displacement of one lattice point resulting in the disruption of translation symmetry. It results in the creation of a pair of adjacent lattice points with incommensurate number of nearest neighbours.  
A disclination, on the other hand, is a distortion of a perfect lattice in which a lattice point gains or loses a few nearest neighbors, breaking the orientational symmetry. Fig. \ref{nernst}(a) shows the schematics of a dislocation which contains two disclinations with five-fold and seven-fold orientational symmetry respectively.

The importance of these defects was recognized by Berezinskii, Kosterlitz, and Thouless (BKT) \cite{bkt1,Kosterlitz_1972,bkt2,bkt3} who studied the topological long-range order in 2D solids and superfluids and the topological solid-liquid phase transition which is applicable to systems such as 2D electron gas, 2D crystals, and neutral superfluids. In a 2D crystal, this transition is manifested by dislocation proliferation due to thermal excitations, resulting in the solid melting into an isotropic liquid. 
By refining and extending the BKT theory, Halperin, Nelson and Young (HNY) predicted the possibility of an intermediate state \cite{2dmelting1, 2dmelting2,PhysRevB.18.2318} between the solid and the liquid, rendering the 2D melting a two-step process. At the BKT temperature, $T_{BKT}$, the solid melts due to thermally excited dislocations into an exotic anisotropic "hexatic fluid", where the system loses its long-range translational order but preserves the quasi-long-range orientational order. In this phase, dislocations can glide along an easy crystalline axis driven by local stress. At higher temperatures, each dislocation unbinds into two independent disclinations and the system transits into an isotropic liquid. This BKTHNY process has been verified experimentally by measurement of shear modulus and colloidal motion in 2D crystal systems \cite{soft, observation, PhysRevLett.58.1200} and theoretically by Monte-Carlo simulations \cite{monte-carlo1, monte-carlo2, PhysRevLett.114.035702}.

Recently, the hexatic fluid phase was observed and studied in another physical system, i.e. the melting process of superconducting vortex lattice (VL) in a weakly disordered amorphous MoGe (a-MoGe) thin film \cite{MoGe1, MoGe2}. Electric transport and scanning tunneling spectroscopy measurements established a phase diagram demonstrating the solid - hexatic - liquid transition. It was found that the hexatic phase is characterized by a very low resistance plateau as a function of magnetic field. While the exact cause for such dissipation is not completely understood, this dissipation was later found to be associated with the extreme sensitivity of the hexatic fluid state when exposed to very small external electromagnetic radiation \cite{PhysRevB.100.214518,DUTTA20201353740}. As is well known, the Nernst effect is a powerful tool to study vortex dynamics in a superconductor\cite{nernst_arnab}. In typical superconductors, the Nernst effect originates almost entirely due to the motion of vortices and the contribution from electrons is negligible. Hence, it seems advantageous to utilize the Nernst effect to study the vortex dynamics in this state.

The Nernst effect refers to the transverse electric field that develops on a conducting sample in the presence of a longitudinal thermal gradient and a perpendicular magnetic field. It has been proven to be a very sensitive experimental tool to probe superconducting vortex motion in the fluctuation regime. A substantial Nernst coefficient, $N=\frac{E_y}{-\nabla_x{T}}$, was measured around $T_C$ in the underdoped regime of high-Tc superconductors \cite{ong, nernst_highTc} and in 2D disordered films \cite{nernst_disorder1, nernst_disorder2, nernst_arnab}.  Fig. \ref{nernst} panels b and c present schematics of the vortex Hall effect and vortex Nernst effect respectively, demonstrating that in the presence of a perpendicular magnetic field, an electrical current or thermal gradient driven vortex motion produces a transverse electric field, $\boldsymbol{E}=\boldsymbol{B}\times \boldsymbol{v}$, where $\boldsymbol{v}$ is the velocity of the mobile vortex. 
Since the Nernst effect in superconductors is governed by vortex motion, it is appealing to ask what would be the Nernst effect of the hexatic phase in which it is dislocations of vortices, rather than vortices themselves, which are mobile. In this paper we present magneto-transport and Nernst effect measurements on weakly disordered a-MoGe thin films. We find a unique Nernst signal sign-reversal in the hexatic phase regime. We compare this to the Hall effect sign-reversal seen in theses systems as well and find them to be unrelated.  We discuss the findings and suggest that they are due to dislocations in the vortex hexatic phase which move from cold to hot.


To sensitively measure the Nernst effect, we adopt an on-board AC heating technique with a heater and two thermometers fabricated on a chip of MEMpax\texttrademark borosilicate glass substrate characterized by low thermal conductivity that made it an excellent choice for Nernst effect measurements (see Fig.\ref{nernst}(d)). The details of the measurement technique are described elsewhere \cite{ac_nernst}. The heater and electrodes are fabricated from $34nm$ thick films of Cr/Au using standard optical lithography techniques. Two thin films of insulating amorphous indium oxide (InO) with thickness 30nm are e-beam evaporated with partial oxygen  pressure of $3 - 4*10^{-5}$ torr, to be used as thermometers that are characterized by excellent temperature sensitivity in the desired measurement range (sheet resistance of 200 - 10 k$\Omega$ at the temperature range of 1-5 K). The sample, a 25-nm thick a-MoGe is deposited between the two thermometers via pulsed laser deposition. In order to easily obtain a temperature gradient, the substrate is suspended from the cold finger with only the edge far from the heater thermally contacted to the thermal bath.  All the measurements were performed on two different samples in a pumped Helium-4 cryostat in the bore of a 9-Tesla superconducting magnet.

\begin{figure}[tb]
\includegraphics[width=0.45\textwidth]{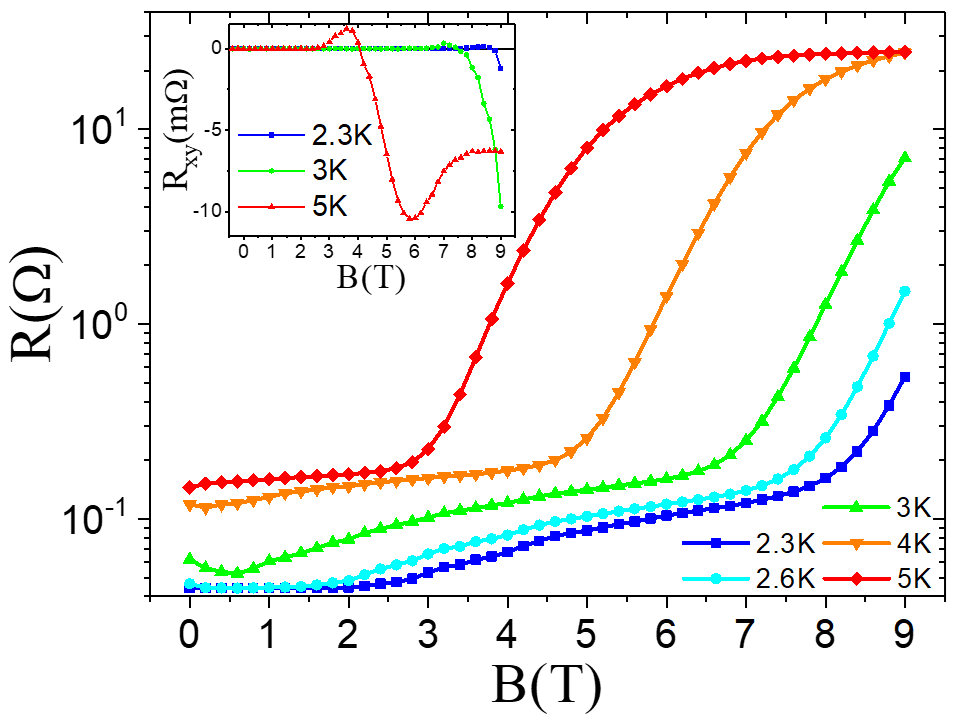}
\caption{(Color online) Magnetoresistance of a 25nm thick a-MoGe film (sample S1) at different temperatures. At a low temperature (2.3 - 3K), system transits from a solid to a liquid through a wide hexatic regime with increasing magnetic field. The pinning is gradually overcome at higher temperatures so that it could not reach solid phase. (Inset) Hall resistance $R_{xy}$ at several temperatures is plotted as a function of magnetic field. The hall resistance vanishes at a low field, and a sign-reversal is captured to happen at the transition between the hexatic phase to a liquid phase. }
\label{mr}
\end{figure}

Fig.\ref{mr} shows the longitudinal and Hall resistance of one of our a-MoGe films as a function of magnetic field for several fixed temperatures. Similar results were obtained for three more films. The results were symmetrized and anti-symmetrized respectively with respect to the magnetic field. At low temperature and magnetic field, the magnetoresistance saturates at 4 m$\Omega$ which sets our zero resistance baseline within our Nernst measurement-setup capability. In this limit the sample is in the vortex solid phase where all vortices are pinned and non-dissipative. As the temperature and magnetic field are increased, the resistance rises and reaches an intermediate plateau, which is identified as the hexatic fluid phase where vortex dislocations are mobile (see \cite{MoGe1}). At an even higher temperature or magnetic field the resistance rises dramatically and saturates at the normal state value. Between these two plateaus (identified as the superconducting vortex liquid regime) the Hall resistance, $\rho_{xy}$ also changes significantely. Deep in the superconducting state it vanishes, above $H_C$ it is constant at its normal state value, and in-between it attaines a positive sign followed by a sign-reversal by which the Hall coefficient remained negative up to the normal state.
 
\begin{figure}[tb]
\includegraphics[width=0.5\textwidth]{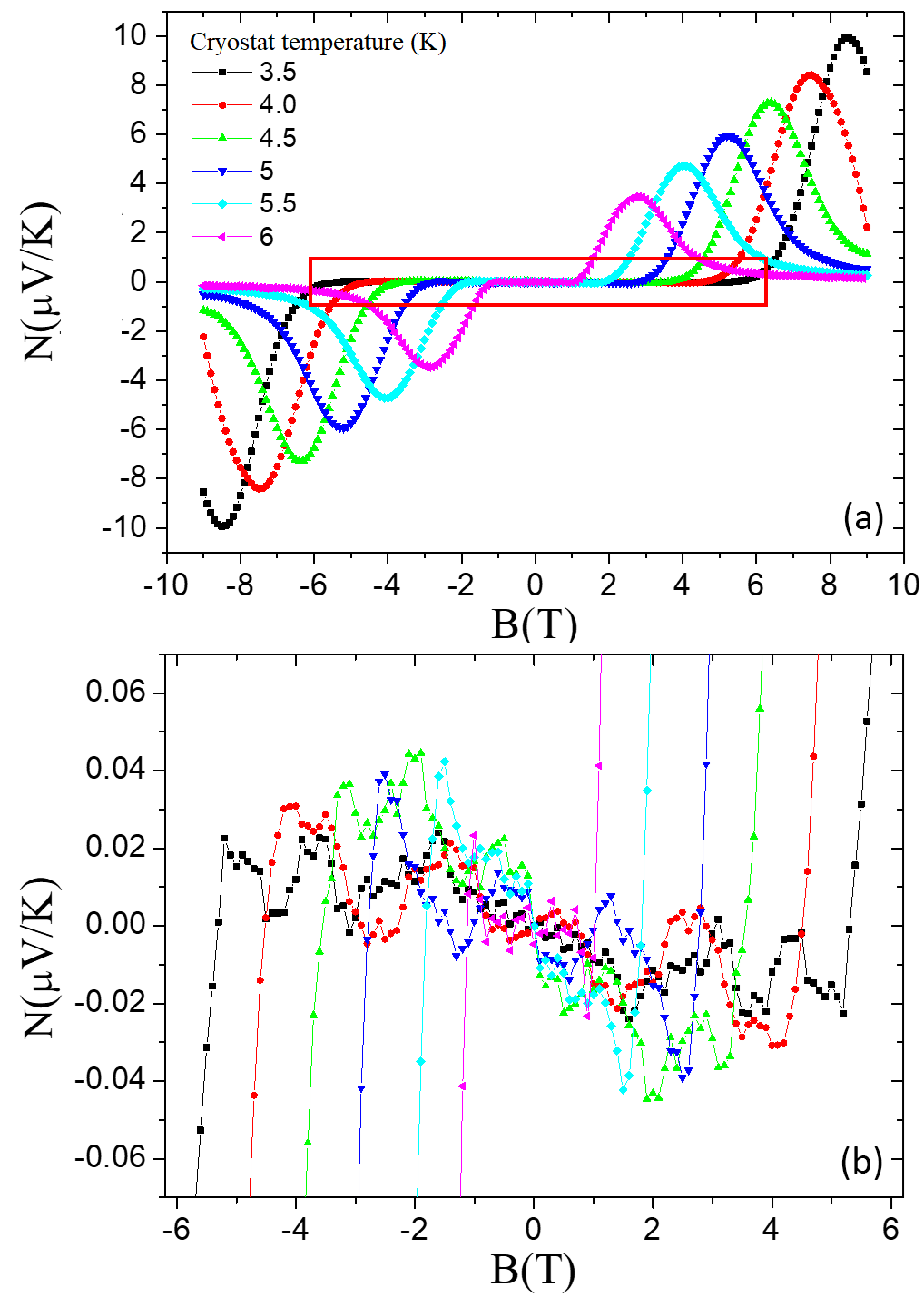}
\caption{(Color online) (a) Nernst signals of sample S2 as a function of magnetic field at different temperatures with AC technique. Due to the vortex pinning, Nernst signals are small at a low field while typical Nernst peaks show up in the superconducting transition regime. (b) by zooming into the regime in the red box in (a), negative Nernst signals are detected. Notably, the slope of the Nernst signal monotonously increases with increasing temperature.}
\label{Nst}
\end{figure}

The Nernst signal $N$ of sample S2 as a function of magnetic field with different temperatures, $T = 3.5 - 6.0$K, is shown in Fig. \ref{Nst}. The data was anti-symmetrized over the field to remove parasitical background voltage. The sample temperature was obtained by averaging the two thermometer temperatures. It is seen that the Nernst signal displays a strong peak (up to 10 $\mu V/K$) at high fields, which corresponds to the vortex motion in the liquid phase. This is a typical result for disordered superconductors \cite{nernst_disorder1, nernst_disorder2, nernst_arnab}.  The Nernst peak shifts towards smaller fields as the temperature is raised due to suppression of superconductivity. 

The main result of this work is shown in Fig. \ref{Nst}b which focuses on the Nernst signal in the low field regime. Here we observe an unusual \textit{negative} Nernst signal with magnitude of  several 10s of $nV/K$. This result is very counter-intuitive implying a vortex diffusion direction against the temperature gradient, from cold to hot. Obviously, such a picture would violate the second law of thermodynamics. With increasing temperature, the field range at which this sign-reversal is observed deceases, but it remains finite for all temperatures below $T_C$.

\begin{figure}[t]
\includegraphics[width=0.48\textwidth]{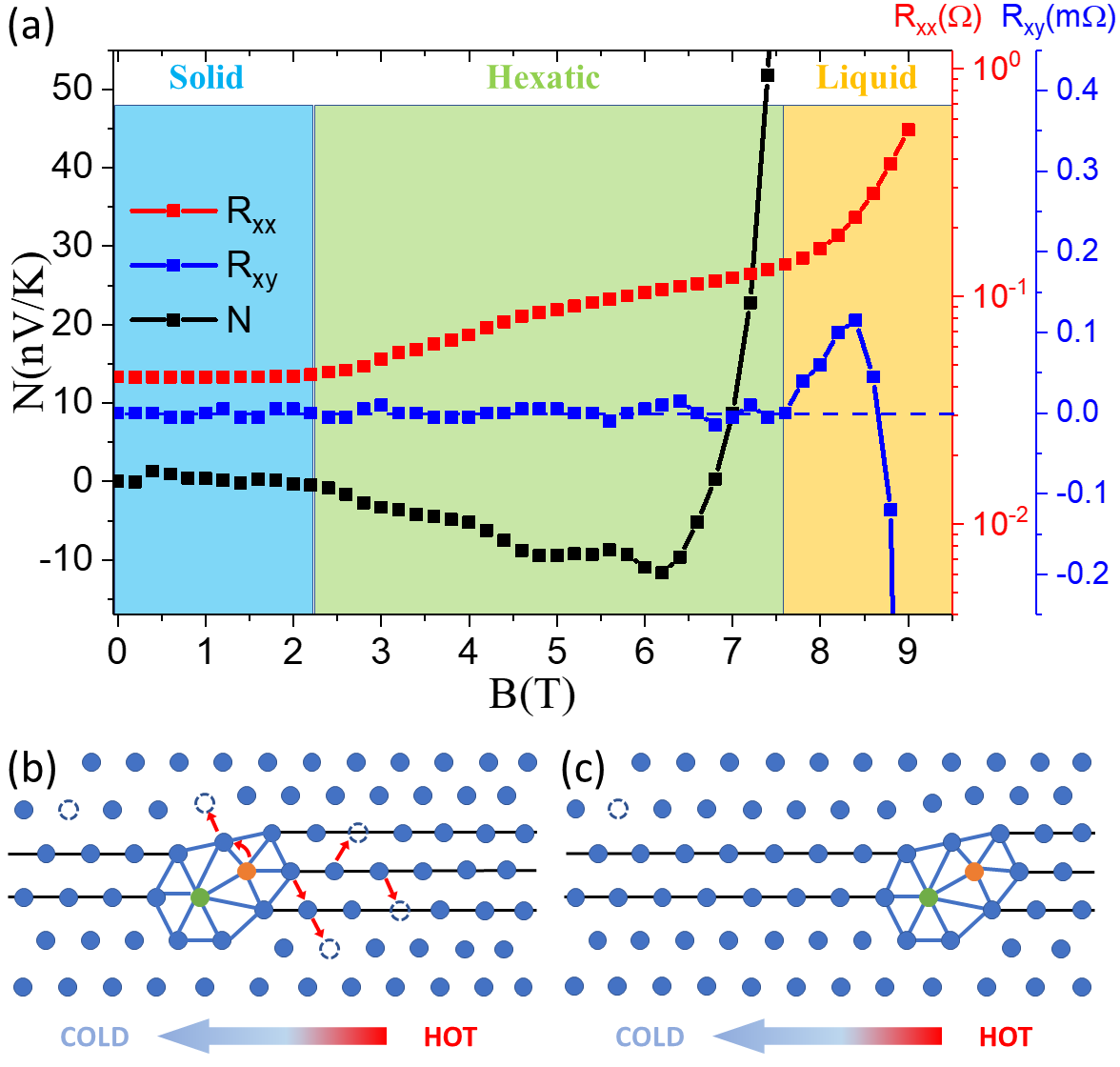} 
\caption{(Color online) (a) Nernst effect $N$, longitudinal resistance $R_{xx}$ and Hall resistance $R_{xy}$ of sample S1 measured at 2.3K are plotted as functions of magnetic field. In the hexatic phase, vortices start to move in the form of dislocations. As a result, $R_{xx}$ is driven in a plateau, where the Nernst effect abnormally tends negative. In the liquid phase, the Nernst effect turns to be positive and shoots up with $R_{xx}$ where $R_{xy}$ exhibits a sign-reversal, the blue dashed line indicates zero resistance in $R_{xy}$. (b) In the hexatic state, vortices near a dislocation diffuse to fill the vacancies (dashed circles), resulting in a dislocation climb into (c) with a net flow of vortices from cold (left) to hot (right)  and leading to a negative Nernst effect. This phenomenon occurs due to the increased concentration of vortex vacancies at higher temperatures. The direction of vortex diffusion is indicated by the red arrows. On the other hand, the position of vacancies accumulated on the dislocation line is occupied by the surrounding vortices due to lattice relaxation, which leads to a perfect lattice forming to the left of dislocation in (c).
}
\label{mrN}
\end{figure}

Fig.  \ref{mrN}a compares $N$, $\rho_{xx}$ and $\rho_{xy}$ of sample S1 at T = 2.3K. Notably, the negative $N$ appears at the same field regime as the plateau of resistance, which is associated with the hexatic state \cite{MoGe1, MoGe2}. At this regime the Hall resistance is immeasurable within our measurement resolution. On the other hand, a measurable Hall sign-reversal effect is seen at higher field which is identified with the transition from hexatic state to vortex liquid, where the Nernst signal is uniquely positive.

A Hall effect sign-reversal, such as that seen in figures \ref{mr} and \ref{mrN} has been widely reported both in conventional type-II superconductors \cite{VANBEELEN1967241,anomaly_mosi} and in high-Tc superconductors \cite{anomaly_htc1, PhysRevB.62.9780, table}. The cause of the Hall sign-reversal in the flux-flow regime of superconductors, unpredictable from the standard model of superconducting vortex by Bardeen and Stephen\cite{BS}, is still under debate and a number of physical origins have been suggested, such as back-flow of vortices in a clean superconductivity limit \cite{backflow1,backflow2,PhysRevLett.68.2524}, interlayer vortex coupling \cite{CuO_plane, anomaly_Moge,vortex_decoupling} and moving vortex charge \cite{assa,vortex_charge,anomaly_htc1}, etc. But each of these has its drawbacks and a complete understanding is still lacking.

Since, within the linear response approximation, N is expressed  as \cite{Behnia_2016}
\begin{equation}
N=\rho_{xx}\alpha_{xy}+\rho_{xy}\alpha_{xx} 
\label{eq_nernst}
\end{equation}
($\alpha_{xx}$ and $\alpha_{xy}$ are the diagonal and off-diagonal elements of the thermoelectric matrix, respectively) a natural cause for the N sign-reversal could be the sign-reversal in $\rho_{xy}.$ Mathematically, there is nothing that prevents the Nernst effect from being negative. However, $\rho_{xx}$ is always positive, and $\alpha_{xy}$, which is interpreted as the entropy carried by each flux quantum ($\alpha_{xy} = \frac{S_{\text{vortex}}}{h/2e}$) \cite{Behnia_2016, Behnia_2023}, is also positive in order to obey the second law of thermodynamics. Therefore, the second term $\rho_{xy} \alpha_{xx}$ is a suspected candidate for the cause of the Nernst sign-reversal. 

In order to explore this possibility we recall that the Seebeck effect $S$ (longitudinal electric field divided by a temperature gradient) is given by
\begin{equation}
 S=-\rho_{xx}\alpha_{xx}+\rho_{xy}\alpha_{xy}
 \label{seebeck}
\end{equation}
We have complemented our Nernst effect measurements by Seebeck experiments (see supplementary) using the same AC technique, in order to extract $\alpha_{xx}$ and $\alpha_{xy}$. We find that in the hexatic phase, has the same order of magnitude as $\alpha_{xy}$. At this regime, $\rho_{xy}$ is at least 3-4 orders of magnitude smaller than  $\rho_{xy}$ (See Fig. \ref{mrN}) thus rendering the right term of Eq. \ref{eq_nernst} negligible compared to the left term. This, together with the fact that the observable Hall effect sign-reversal occurs at high fields, in a regime where the Nernst coefficient is entirely positive, rules out the Hall effect as the origin for the Nernst sign-reversal. 

The fact that the negative Nernst signal regime coincides with the hexatic state MR plateau, implies that the two are closely related. A sign change in the Nernst signal can imply one of two scenarios. Firstly, it is possible that vortex contribution to the Nernst signal is totally suppressed in the hexatic state, and what we measure is the electronic contribution to the Nernst effect originating in the leads. Secondly, it is also possible that the effective direction of motion of the vortices in a thermal gradient has been reversed due to the hexatic order \cite{PhysRevB.81.060508,PhysRevB.83.174502}. To rule out the first possibility, we have conducted experiments to measure the Nernst effect of the leads alone under conditions identical to those used while measuring MoGe. The results (see supplementary) indicate that the Nernst effect of the leads are too small to measure even for our sensitive experimental setup. The second scenario may be realized if instead of vortices, it is the vacancies or voids in the otherwise perfect vortex lattice that are on the move from the region of high temperature to the region of low temperature. A mechanism that explains such motion of vacancies is "dislocation climb" \cite{PhysRevB.25.579,Nabarro,hull2011introduction}, which describes the movement of a dislocation mediated by the thermal motion of vacancies. In the hexatic state, the vortices are still pinned to their locations, but defects, mostly dislocations, are free to move within the lattice. The motion of these dislocations constitutes the major dissipation mechanism in the hexatic state \cite{PhysRevLett.69.2709}. Naturally, movement of these dislocations also play an important role in the origin of the Nernst effect. In dislocation climb, vortices from the dislocation overcome the potential barriers due to thermal energy and migrate to fill nearby thermally activated vortex vacancies to attain a lower energy state as schematically described in Fig. \ref{mrN}b,c.  In the presence of a non-equilibrium distribution of vacancies generated by an external temperature gradient (i.e. an increased concentration of vortex vacancies at higher temperatures), vortex dislocations in the hexatic phase move preferably in the  direction opposite to the applied temperature gradient, resulting in a net flow of vortices from cold to hot and a negative Nernst effect, while the overall vortex lattice remains nearly frozen. Meanwhile, a line of vortex vacancies are accumulated at the tail of the moving dislocation. Upon immediate relaxation of VL, vacancies are occupied by the vortices which were squeezed out by the original dislocation. A perfect lattice structure is then formed on the cold side of the dislocation after the climb takes place as sketched in Fig. \ref{mrN}c. This scenario is consistent with the increasing slope of the Nernst coefficient  with increasing temperature (shown in Fig. \ref{Nst}b), since the density of both vortex dislocations and vacancies are expected to increase with increasing temperature. Hence, the motion of dislocations and the arrangement of the VL play a crucial role in the emergence of the negative Nernst effect in the hexatic phase.

In conclusion, Nernst sign-reversal and Hall sign-reversal were both observed in weakly disordered superconducting a-MoGe films. While the Hall sign-reversal happens in the isotropic vortex fluid phase, the negative Nernst signal and its sign-reversal occurs in the hexatic phase. The unusual negative Nernst effect is attributed to the motion of the vortex dislocations and their anomalous diffusion against the temperature gradient. One possibility for such an effect is via dislocation-climb through thermally activated vortex vacancies of non-equilibrium concentration under a temperature gradient, which leads to a net backflow of vortices and a negative Nernst effect. The 2D dislocation climb has been experimentally demonstrated in 2D transition metal dichalcogenides \cite{climb_TMDC} in a study of migration of dopant atoms and indirectly evidenced in some colloidal systems \cite{PhysRevLett.73.3113}, but has not yet been detected in vortex lattice systems. Therefore, further experimental and theoretical studies are needed to confirm the  and fully understand this proposed mechanism.

\begin{acknowledgments}
The authors would like to thank D.P. Arovas, A. Auerbach, E. Shimshoni and V. Vinokour  for valuable discussions and I. Volotsenko and G. N. Daptary for technical help. This research was supported by The Israel Science Foundation, ISF grants no. 3192/19 and 1499/21.
\end{acknowledgments}

\bibliography{references}

\newpage
\begin{center}
    \textbf{ \large Supplemental Material}
\end{center}

\section{Seebeck effect and Calculations of $\alpha_{xx}$ and $\alpha_{xy}$ } 

We measured Seebeck coefficient, S, of sample S1 as a function of magnetic field using the same AC method as that for Nernst measurements. At high fields Seebeck signals are vanishingly small compared to the Nernst signals due to particle hole symmetry. However, in the hexatic state, S and N are of the same order of magnitude. 

\begin{figure}[htb]
\centering \includegraphics  [width=9cm,height=!]{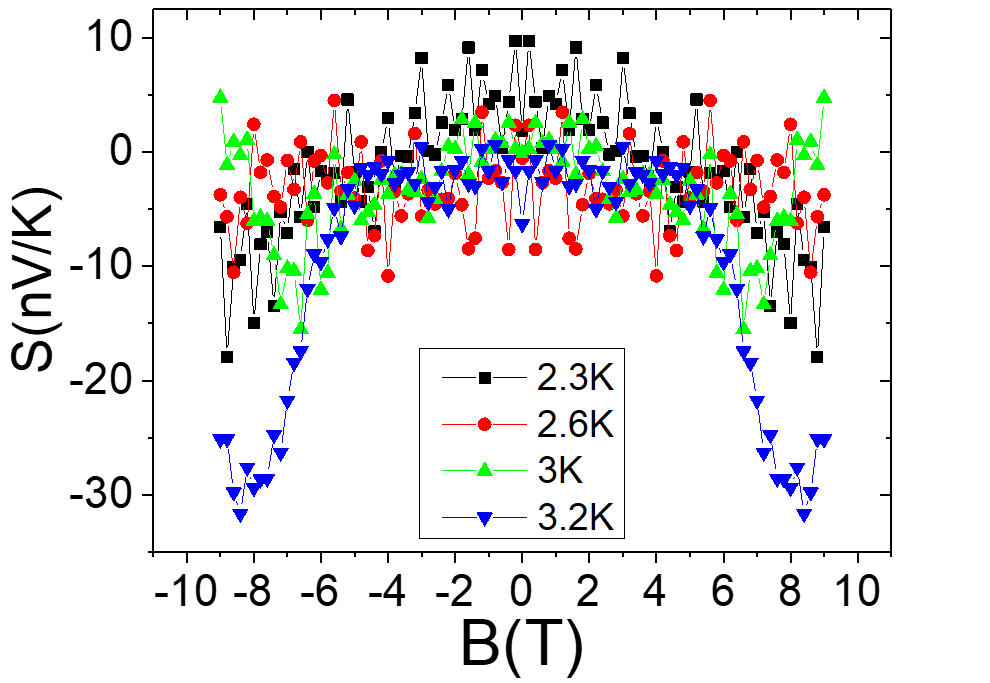}
\caption{\label{seeabeck}Seebeck coefficient measured as a function of magnetic field at various temperatures. All the resulting curves were symmetrized with respect to magnetic field to remove signals caused by misalignment of the leads.
 }

\end{figure}

From the results of Nernst coefficients and Seebeck coefficients, We extract the thermoelectric tensor elements $\alpha_{xx}$ and $\alpha_{xy}$ by definition as follows 

\begin{equation}
    \alpha_{xx} = \frac{\rho_{xy} N-\rho_{xx} S}{\rho_{xx}^2+\rho_{xy}^2}
\end{equation}
and 
\begin{equation}
    \alpha_{xy} = \frac{\rho_{xx} N+\rho_{xy} S}{\rho_{xx}^2+\rho_{xy}^2}
\end{equation}

\begin{figure} [htb]
\centering \includegraphics  [width=8.5cm,height=!]{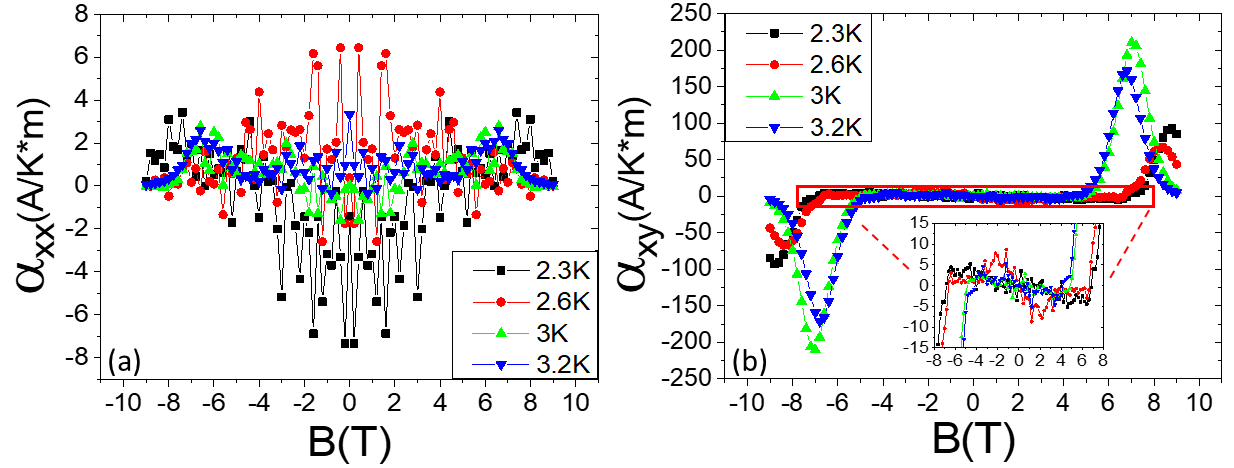}
\caption{\label{alpha} (a) $\alpha_{xx}$ and (b) $\alpha_{xy}$ extracted as a function of magnetic field from the measurements of Nernst effect and Seebeck effect at various temperatures. The inset of (b) zooms in on the $\alpha_{xy}$ values in the red rectangle, where a sign-reversal in the Nernst effect was found, and shows a similar sign-reversal in $\alpha_{xy}$.
 }
\end{figure}

Fig.\ref{alpha} (a) and (b) present the calculated results of $\alpha_{xx}$ and $\alpha_{xy}$. Notably, $\alpha_{xy}$ is negative in the same regime where the negative Nernst effect appears. This phenomenon cannot be explained without assuming there is anomalous diffusion of vortices due to the interaction between vortex dislocations and vacancies

\section{Nernst Signal Contributions from Chromium and Gold (Cr/Au) Contacts} 
\begin{figure}[htb]
\centering \includegraphics  [width=8cm,height=!]{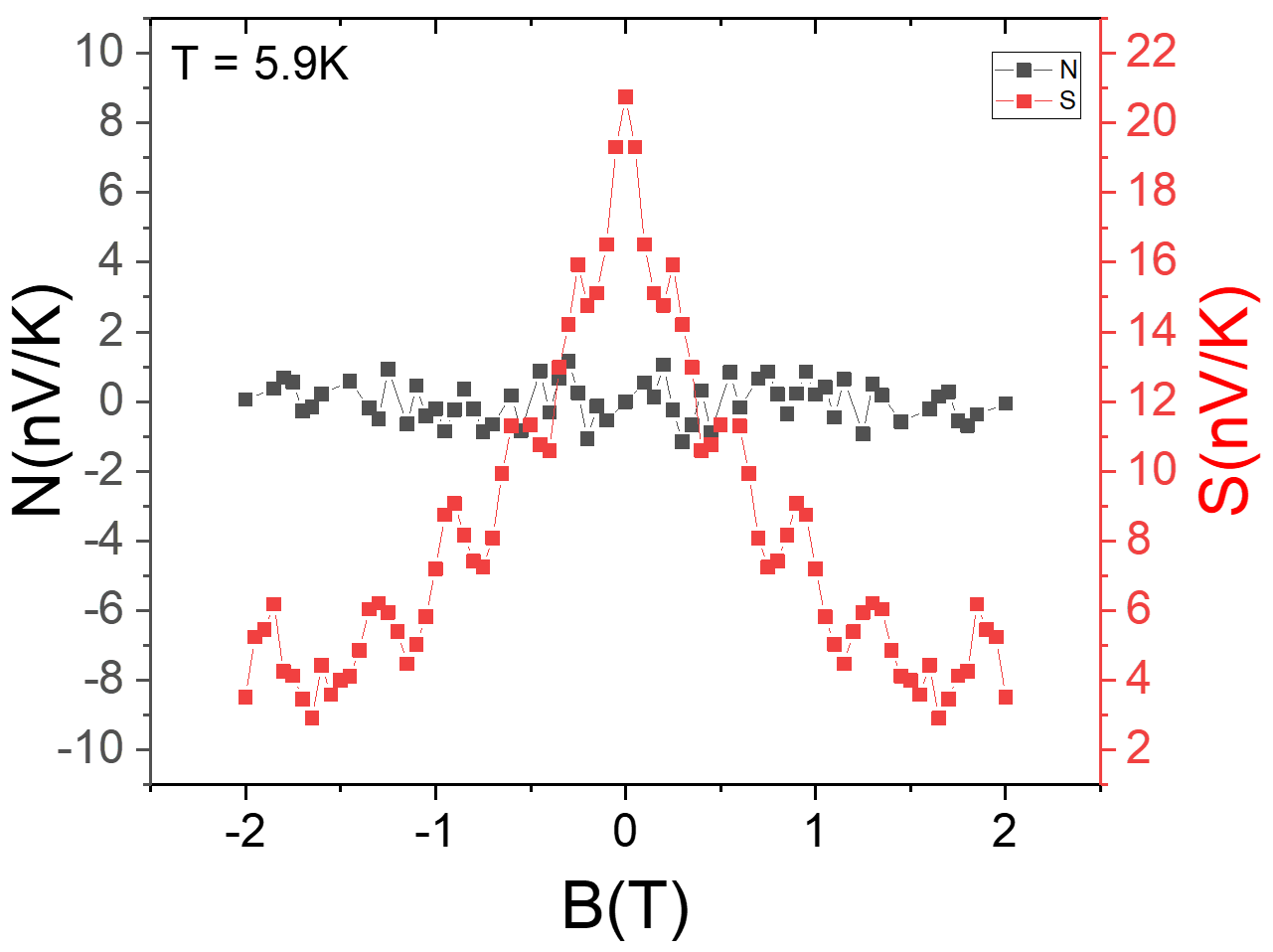}
\caption{\label{CrAu} Nernst (black) and Seebeck (red) coefficients of a Cr/Au film measured at 5.9K by our AC technique.
 }

\end{figure}

To eliminate a potential Nernst contribution from the Cr/Au contacts in our MoGe sample, we substituted the MoGe film with a film of 4nm thick chromium and 30nm thick gold, identical to our contact pads, while maintaining the thermometers and a heater using the same procedure as described in the main text. The results for Nernst and Seebeck effect measurements conducted at low temperatures are presented in Fig. \ref{CrAu}.

While we observe a finite Seebeck coefficient for the Cr/Au film, the Nernst coefficient remained below our measurement sensitivity of $\approx {1nV/K}$. This provided an upper bound for the Nernst coefficient of the contacts in our experimental setup. Furthermore, we note that for sample S1 described in the main text, the width of the Cr/Au contact was $w = 20\mu\text{m}$ while in our test-case the distance between the Nernst voltage leads was 500 $\mu$m. Taking into account that the typical magnitude of the negative Nernst coefficient for MoGe is approximately 20 nV/K, we estimate the upper bound ratio of the Nernst signal contribution from a contact relative to the signal from the MoGe film to be at most:

\begin{equation}
\begin{aligned}
    \frac{V_{contact}}{V_{MoGe}} = \frac{N_{contact}}{N_{MoGe}} \times \frac{w}{w_{MoGe}} \\
    \le \frac{1\rm{nV/K}}{20\rm{nV/K}} \times \frac{20\mu m}{500 \mu m} = 0.002
\end{aligned}
\end{equation}

Hence, the contact-generated signal is at most three orders of magnitude smaller than the magnitude of the negative Nernst signals reported in the manuscript.

\section{Procedures of Curve Symmetrization and Antisymmetrization}
In this letter, we present the experimental studies of several physical quantities, including magnetoresistance (MR), Hall resistance (HR), Nernst coefficient and Seebeck coefficient, as functions of magnetic field in superconducting MoGe. Since it is a non-magnetic material, MR and Seebeck coefficient are expected to be symmetric with respect to the applied magnetic field, while HR and the Nernst coefficient should be antisymmetric. Naturally, in real measurements, there are unavoidable (anti)symmetric biases due to various factors, such as misalignment of contacts and stray voltages. Hence, we perform (anti)symmetrization to remove the parasitic biases.

\begin{figure}[h]
\centering \includegraphics  [width=8cm,height=!]{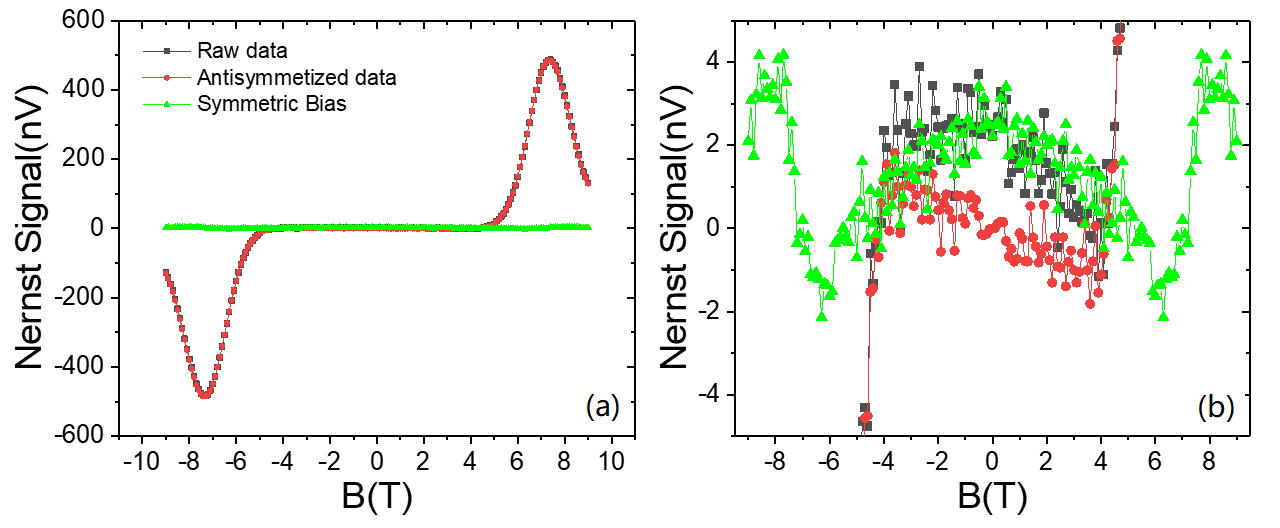}
\caption{\label{antisym} (a) Nernst signals for a 30nm-thick MoGe (Sample S3) measured at 3.5K from -9 to 9T, raw data (black) is decomposed into an antisymmetrized curve (red) and a symmetric bias (green). (b) zoom into the the regime of negative Nernst signals.
 }
\end{figure}

Based on the mathematical fact that an arbitrary function of magnetic field, denoted as $V(B)$, can be decomposed into a symmetric function, $V_{sym}(B)$, and an antisymmetric function, $V_{asym}(B)$, where
\begin{equation}
\label{sym}
V_{sym}(B) = \frac{V(B) + V(-B)}{2}
\end{equation}
and
\begin{equation}
\label{asym}
V_{asym}(B) = \frac{V(B) - V(-B)}{2}
\end{equation}

we proceed with the following detailed procedure. At fixed temperatures, we conducted measurements of MR, HR, Nernst and Seebeck coefficients at intervals of 0.1T in the magnetic field range of -9 to 9T. To eliminate any residual background voltage, we performed numerical (anti)symmetrization  of the raw data. An example of Nernst signals before and after antisymmetrization is shown in Fig. \ref{antisym}.

\end{document}